# Quality assurance test and Failure Analysis of SiPM Arrays of GECAM Satellites

D.L. Zhang[*1], M. Gao[1], X.L. Sun[*2], X.Q. Li[*3], Z.H. An[*4], X.Y. Wen[1], C. Cai[1], Z. Chang[1], G. Chen[1], C. Chen[1], Y.Y. Du[1], R. Gao[1], K. Gong[1], D.Y. Guo[1], J.J. He[1], D.J. Hou[1], Y.G. Li[1], C.Y. Li[1], G. Li[1], L. Li[1], X.F. Li[1], M.S. Li[1], X.H. Liang[1], X.J. Liu[1], Y.Q. Liu[1], F.J. Lu[1], H. Lu[1], B. Meng[1], W.X. Peng[1], F. Shi[1], H. Wang[1], J.Z. Wang[1], Y.S. Wang[1], H.Z. Wang[1], X. Wen[1], S. Xiao[1], S.L. Xiong[1], Y.B. Xu[1], Y.P. Xu[1], S. Yang[1], J.W. Yang[1], Fan. Zhang[1], S.N. Zhang[1], C.Y. Zhang[1], C.M. Zhang[1], Fei Zhang[1], X.Y. Zhao[1], X. Zhou[1]

[1] *Institute of High Energy Physics, CAS, Beijing 100049, China;*
* E-mail: zhangdl@ihep.ac.cn

[2] *Institute of High Energy Physics, CAS, Beijing 100049, China;*
* E-mail: sunxl@ihep.ac.cn

[3] *Institute of High Energy Physics, CAS, Beijing 100049, China;*
* E-mail: lixq@ihep.ac.cn

[4] *Institute of High Energy Physics, CAS, Beijing 100049, China;*
* E-mail: anzh@ihep.ac.cn

Abstract: The Gravitational wave high-energy Electromagnetic Counterpart All-sky Monitor (GECAM) satellite consists of two small satellites. Each GECAM payload contains 25 gamma ray detectors (GRD) and 8 charged particle detectors (CPD). GRD is the main detector which can detect gamma-rays and particles and localize the Gamma-Ray Bursts (GRB),while CPD is used to help GRD to discriminate gamma-ray bursts and charged particle bursts. The GRD makes use of lanthanum bromide ($LaBr_3$) crystal readout by SiPM. As the all available SiPM devices belong to commercial grade, quality assurance tests need to be performed in accordance with the aerospace specifications. In this paper, we present the results of quality assurance tests, especially a detailed mechanism analysis of failed devices during the development of GECAM. This paper also summarizes the application experience of commercial-grade SiPM devices in aerospace payloads, and provides suggestions for forthcoming SiPM space applications.

# 1. Introduction

After the first detection of gravitational wave in 2016, the "Gravitational Wave Electromagnetic Counterpart All-sky Monitor (GECAM)" project was proposed to monitor the gravitational waves related GRBs, and guide subsequent observations in other wavelength of EM.

On August 17, 2017, LIGO and Virgo jointly observed a gravitational wave (GW170817) originating from a merger of two neutron stars [1]. At 1.7 seconds after the merger event, the Gamma-ray Burst Monitor (GBM) onboard Fermi [2] independently observed a gamma-ray burst (GRB 170817A) related to the gravitational wave event. This gravitational wave event marks the beginning of the "multi-messenger astronomy" era[3], and triggered many researches in the field of detecting electromagnetic counterparts of gravitational waves.

GECAM is composed of two satellites with orbit height of 600 kilometers and inclination angle of 29°. Each GECAM satellite contains 25 gamma ray detectors (GRD) [4] (Figure 1) and 8 charged particle detectors (CPD)[5][6]. GRD is the main detector for measuring the gamma-rays and particles. The designed energy range is 8keV- 2MeV and tested result is 5.9 keV-4.3 MeV; CPD is used to identify gamma-ray bursts and charged particle bursts. localization algorithm based on the relative count rates and energy spectrum analysis of GRDs[7].

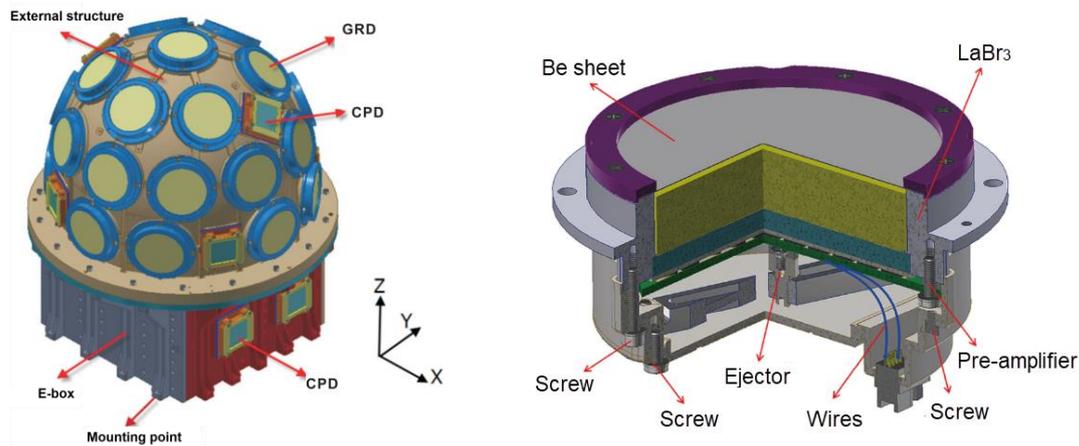

Figure 1 Schematic diagram of GECAM structure. Left: Installation drawing of GECAM GRD detector Right: GRD structure diagram.

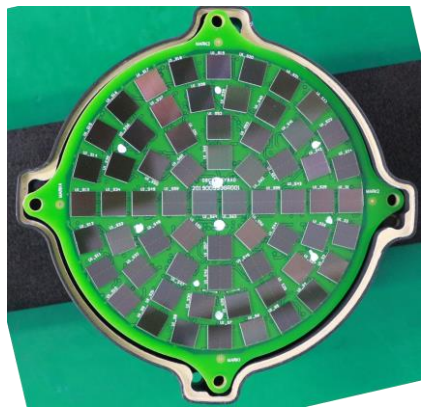

Figure 2 SiPM array of GRD

GRD is primarily composed of lanthanum bromide ($LaBr_3$) crystals[8][9] and a pre-amplifier board. In the front of the pre-amplifier board there is a SiPM array. GRD uses a novel device SiPM for large-size crystals readout. SiPM is widely used in the field of nuclear medicine imaging [10]. Due to its compact size, low operating voltage, and high quantum efficiency, SiPM has gradually been used in nuclear physics and related fields in recent years. The in-flight calibration detector of the Hard X-ray Modulation Telescope (HXMT) [11] apply silicon photomultiplier tube (SiPM) for the first time in China. The U.S. Naval Laboratory used SiPM as the $SrI_2$ scintillator signal readout in the SIRI-1 payload [12] launched in 2018. The detectors of Tsinghua University's "Gamma Ray Integrated Detectors" (GRID) CubeSat project [13] are mainly composed of a novel scintillator GAGG and a SiPM array. The above projects are all small-scale SiPM applications, and the quantity is only a few dozen pieces. Each SiPM array of the GECAM project consists of 64 SiPM chips (Figure 2), the quality grade of Model SensL MicroFJ-60035-TVS is commercial grade. In order to avoid the failure of the entire SiPM array due to the failure of single SiPM, SiPM's power supply network is divided into two parts that can switch each part separately. There are 3200 SiPM chips on GRDs and such a large-scale SiPM space application requires an efficient and reliable aging test to ensure the normal operation of the detector in orbit. Therefore, the quality assurance test is to test the entire pre-amplifier board as a whole. If one SiPM fails, the entire pre-amplifier board will be discarded. As one piece of SiPM failed in the joint payload test, a detailed failure analysis was carried out.

## 2. Quality assurance test of SiPM arrays and pre-amplifier

As the SiPM used by GECAM is a commercial grade device (MICROFJ-60035-TSV-TR), it must be screened through aging. The objective of the quality assurance test is to identify the potential defective components as early as possible. The pre-amplifier board is tested as a whole. Once a chip of the SiPM fails or the nonlinearity of the amplifier does not meet the requirements of ＜3%, the whole pre-amplifier board will be discarded. The final assembled 50 SiPM arrays for the GECAM payload is select according to lower SiPM array dark current of ＜110 μA that means a lower electronic noise and better low energy detection range.

In the 79 SiPM arrays, the surface glass layers of some SiPMs are broken by accidental impact during assembly. As a result, the micro-cells of the SiPM is damaged. The cathnode pin of two SiPM arrays are short-circuited to the ground, and the leakage current of one SiPM array is larger than 200 μA. Excluding the three pre-amplifier boards mentioned above, 76 boards participated in the aging test. The aging tests comprise mechanical tests and the atmospheric temperature cycle tests according to Chinese national military standard. The screening test of pre-amplifier board includes three items: LED array test, integral nonlinearity test of the pre-amplifier and dark current test.

In order to quickly judge whether the SiPM chip has false soldering or short circuit fault of SiPM microcell caused by surface glass layer fracture, a LED array is designed according to the SiPM arrangement in Figure 3. After the LED array, collimator and SiPM array are completely assembled, Microcontroller Unit(MCU) and analog switch controls the led array to emit light in the order of S1-S64, with an interval of 400 μS. The right panel of Figure 3 shows the signal waveforms of each cell of SiPM array in turn. The output pulse amplitude of GRD preamplifier is different due to the difference of luminous intensity between LEDs. If the signal output of a SiPM chip is lower than the threshold of 200 mV, the whole SiPM array will be excluded. The test results of LED array can quickly determine the location of fault SiPM, which is convenient for further failure analysis. The data sheet shows that there is an optical uniformity of ±9% among the SiPM[14]. However, because the 64 signal outputs of SiPM array are linked together, there is no strict requirement for the uniformity of SiPMs. The position response uniformity of the whole GRD is within 2% which is reported in a dedicated paper on the preamplifier of GRD[15].

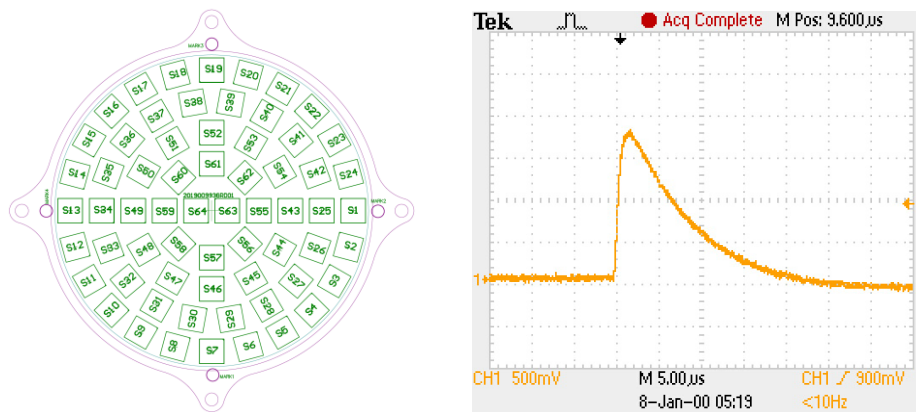

Figure 3 LED array test. Left: SiPM array number. Right: Oscilloscope waveform of pre-amplifier output signal.

The performance of pre-amplifier is tested by Integral Non-Linearity (INL). The INL (σ) is defined as function(1.1). ΔYmax is the maximum deviation of the tested output voltage from the linear fitting results. Y is the full scale value. A signal generator is used to input the pulse waveform, and data acquisition system record the output signal amplitude for data analysis. Figure 4 left panel shows one set of the linearity test results of high gain channel and low gain channel. The fitting parameter p1 stands for the gain of pre-amplifier. The central INL value of 76 GRD pre-amplifier board are 0.44 % and 0.59 % for the initial test and after screening test. In Figure 4 right, the INL is basically unchanged after all aging test and meets the requirements of ＜ 3%. The relative INL residual from the mean INL is also plotted in Figure 4 right. The residual more than 50% is come from experimental measurement statistical error. The corresponding energy ranges for high gain and low gain channel are 5.9-350 keV and 80 keV-4.3 MeV.

$$\sigma = \left|\frac{\Delta Y_{max}}{Y}\right| \times 100\% \quad (1.1)$$

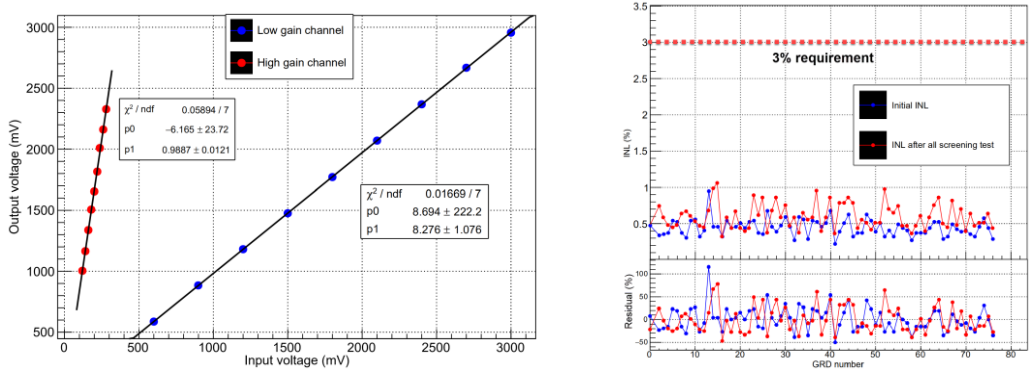

Figure 4 Left: Pre-amplifier linearity of high gain and low gain channels. Right: Integral nonlinearity (INL) of all GRD pre-amplifiers.

The initial tested dark current distribution of the SiPM array at a temperature of 25°C with an operating voltage of 28V can be fitted by Landau function as show in Figure 5 left. The 92.7% unilateral confidence interval of fitted Landau distribution function is 0-200 μA. So SiPM array with current greater than 200 μA are regarded as unqualified with a screening rate around 7%. The most possible values(MPV) of all SiPM arrays of initial test、after mechanical test and after cycling test are 85.6 μA、104.4 μA、121.3 μA respectively. The dark current has a slight increase after the mechanical and temperature cycling test for SiPM arrays and the relative deviations are plotted in Figure 5 right. SiPM array of dark current variation more than 50% can only use as ground test sample. Such SiPM array has some defects before or after the aging test. The SiPM arrays are classified by dark current. 50 SiPM arrays of lower dark current is selected to assembly with better LaBr$_3$ crystals to achieve a detection range requirement lower than 6 keV.

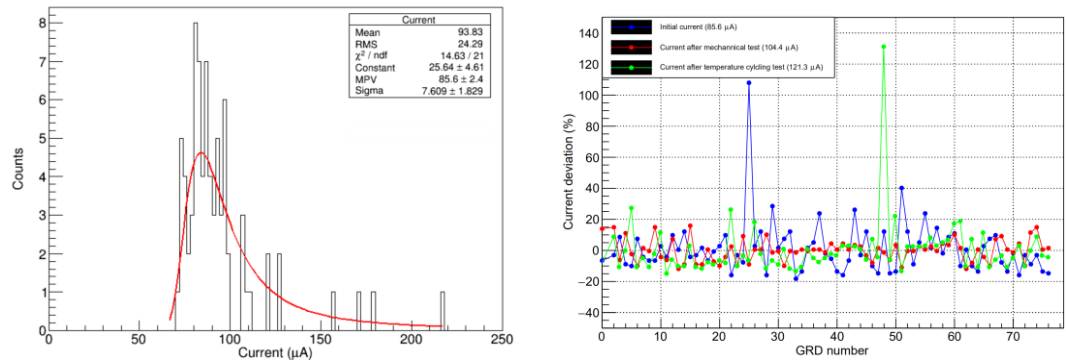

Figure 5 Left: Distribution of SiPM array dark current before aging. Right: Dark current deviation after each aging test. The deviation is defined as the relative deviation of the tested dark current from fitted most possible value(MPV) of dark current among all the SiPM arrays.

SiPM noise is related with dark current and has a total impact on the GRD low energy gamma-ray detection. The current in-flight results show that the dark current of SiPM array increases with time due to radiation damage[16]. The GRD noise threshold is three times the sigma value of the baseline fluctuation. Then the ADC value of threshold is convert to energy according to the in-flight calibration and temperature correction. The in-flight results in left panel of Figure 6 indicate that GRD noise threshold is proportion with SiPM array dark current. At -20 ℃, the low energy detection limit is 6 keV with a dark current of 100 μA. Then the energy threshold increases by 0.26 keV per 100 μA.

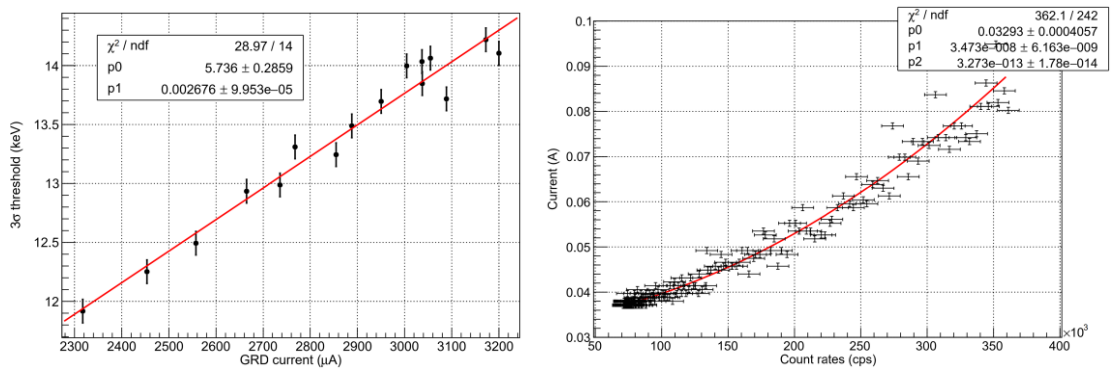

Figure 6 Impact of SiPM array performance changes. Left: GRD threshold vs SiPM array dark current. Right: Dark current increase with count rates.

SiPM dark current is also increased with detection count rates. The right panel of Figure 6 shows an in-flight case of total SiPM dark current increase. In this case, all the SiPM arrays didn't shut down in time at the edge of South Atlantic Anomaly (SAA) area. The high charged particle flux in SAA region leads to the sharp rise of count rates and current. The maximum power supply capacity of the load is 0.2A. If the SiPM array can't be shut down during the whole time of SAA area, the overload of current will have serious damages on SiPM arrays, power module and temperature control system. This risk is mainly due to control command error or the short circuit of SiPM.

However, during the quality assurance, the SiPM arrays are not tested with the whole payload data acquisition system, some test items is not included. Therefore, it was still found that the power supply of one SiPM can't switch off during the whole payload test. This SiPM failure analysis is discussed it the next section.

## 3. SiPM failure analysis

During the joint test of the GECAM payload, it was found that the power supply of No. Z01-40 GRD was uncontrollable. Afterwards, it was disassembled for inspection at Shandong Institute of aerospace electronic technology. As shown in Figure 7, the resistance between the 29.5V reference ground network T1 of the GRD and the power ground network was 68.5 Ω. After removing the R55 with a resistance value of 51 Ω, the resistance of test point T1 to ground was high resistance state, and the resistance of P1 to ground RH was 17.5 Ω(Figure 7 right). The resistance of point T1 to ground was R55 (51 Ω) plus RH (17.5 Ω), which is consistent with the T1 resistance to ground of 68.5 Ω measured before R55 was removed.

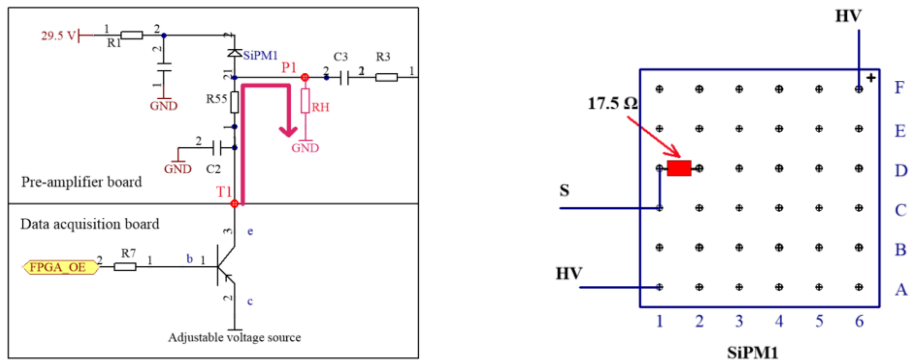

Figure 7 Left: Explanation of SiPM failure status. Right: Resistance value of SiPM failure pins

When the power ground network of the SiPM output P1 generated a leakage channel RH (a dozen ohms), an additional fault path of 29.5V reference ground (T1) to the power ground (blue line in Figure 7 left) would be generated. The fault path was: The current flowed from 29.5V through SiPM and equivalent fault resistance RH, and then to the power ground. The fault path made it impossible for the triode of the data acquisition board (connected to point T1) to control the switching of the power supply ground from point T1.

The fault SiPM was dismantled and delivered to the Space Center of the Chinese Academy of Sciences for device failure analysis. As shown in Figure 8, a microscope was used to inspect the SiPM surface. It was found that the front glass was not broken, and the FR4 layer on the back was visually intact.

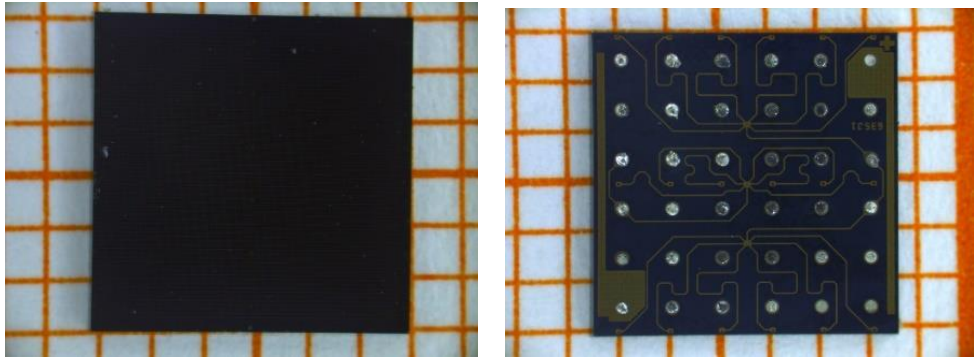

Figure 8 Left: Photo of the front glass body of the malfunctioning SiPM. Right: Photo of the back of the malfunctioning SiPM.

As shown in Figure 9 left, the SiPM device includes 5 layers: silicon substrate, passivation layer ($SiO_2$), titanium nitride barrier layer, copper wiring layer and FR4 barrier layer (transparent) from top to bottom. The two channels (A and B in Figure 9 left) with reduced resistance are located on the surface of the passivation layer. The transparent FR4 barrier layer above the channel is intact. As the scratch position is below the FR4 barrier layer, it is determined that the scratch occurred during the SiPM production process.

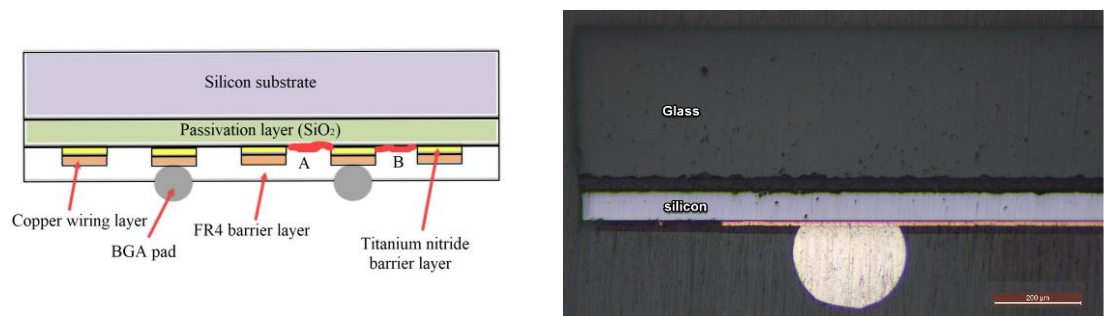

Figure 9 Left: structure diagram of the longitudinal section of SiPM component. Right: metallographic sectioning photo of SiPM

As shown in Figure 10 left, a clearer photo of SiPM was taken by metallurgical microscope, and there are two conductive channels between the D2 pad and the adjacent wire that is connected with D1 pad. The FR4 barrier layer was further rub down to completely expose the conductive path. As shown in Figure 10, after cutting off channel A on the left, the measured resistance of D1 and D2 changed from the original 17.5 Ω and 1.147 kΩ respectively. After cutting off the right channel B, there was a high resistance between D1 and D2. Therefore, it can be judged that the D1 and D2 leakage channels of SiPM are caused by the A and B channels.

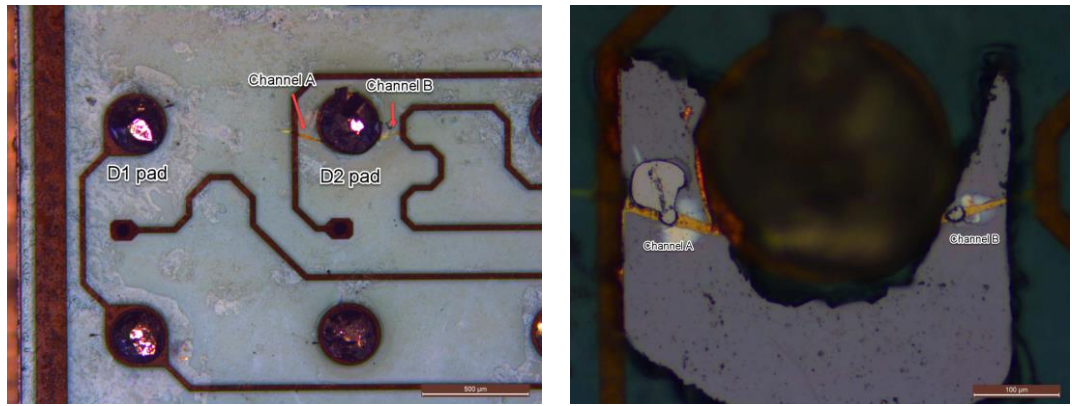

Figure 10 Left: Photomicrograph of faulty pad D2. Right: Photomicrograph after cutting off the conductive path Component analysis for Channel B.

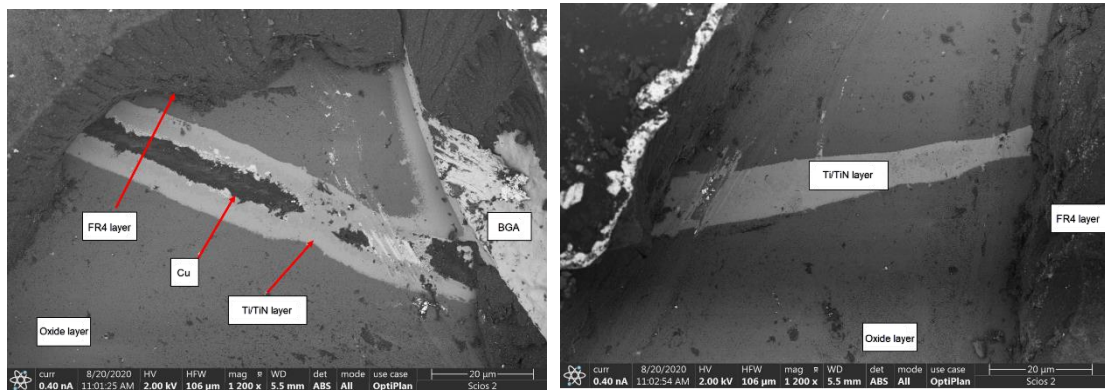

Figure 11 Electron microphotograph. Left: Channel A. Right: Channel B. Component analysis for Channel B.

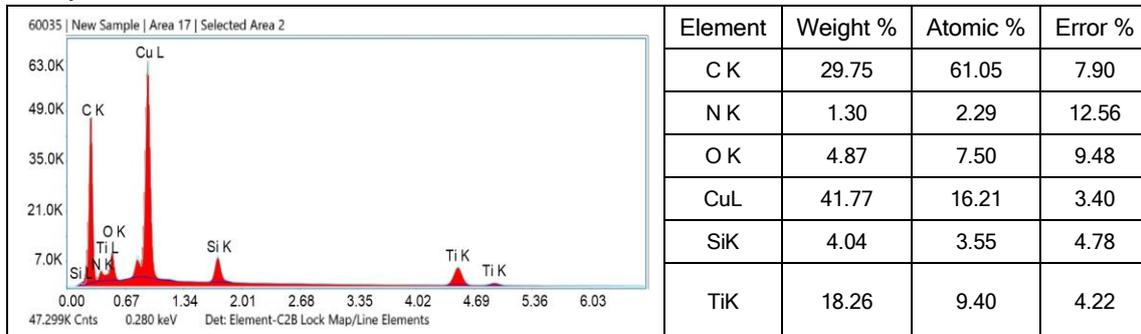

| Element | Weight % | Atomic % | Error % |
| --- | --- | --- | --- |
| C K | 29.75 | 61.05 | 7.90 |
| N K | 1.30 | 2.29 | 12.56 |
| O K | 4.87 | 7.50 | 9.48 |
| CuL | 41.77 | 16.21 | 3.40 |
| SiK | 4.04 | 3.55 | 4.78 |
| TiK | 18.26 | 9.40 | 4.22 |

Figure 12 Component analysis for A channel. Component analysis for Channel B.

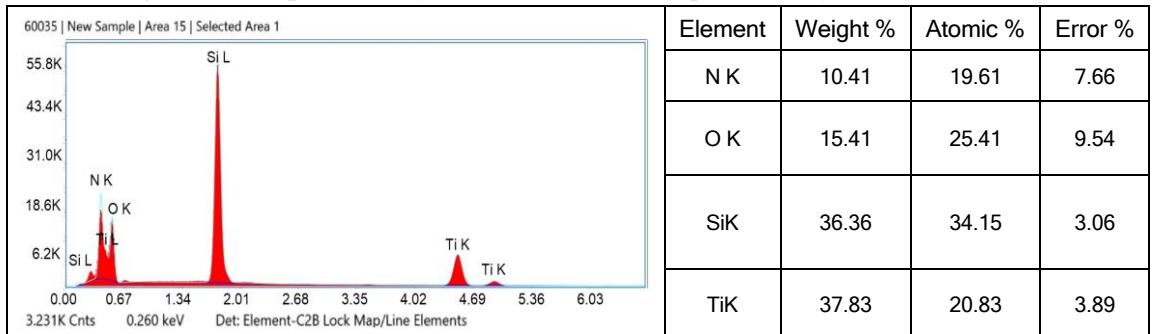

| Element | Weight % | Atomic % | Error % |
| --- | --- | --- | --- |
| N K | 10.41 | 19.61 | 7.66 |
| O K | 15.41 | 25.41 | 9.54 |
| SiK | 36.36 | 34.15 | 3.06 |
| TiK | 37.83 | 20.83 | 3.89 |

Figure 13 Component analysis for channel B. Component analysis for Channel B.

Figure 11 shows the scanning electron tunneling microscope photo of channels A and channel B. In Figure 12 and Figure 13, the elements of each material at channel A and B were analyzed. The oxide layer was silicon dioxide ($SiO_2$), the leakage channel A was titanium (Ti) or titanium nitride (TiN), and copper (Cu) was attached to the top of the film. For channel B, the leakage channel was composed of TiN. This scratch is the main reason for the fault of SiPM and the etching residue of the titanium barrier layer is the fundamental cause. Based on the above analysis, the titanium/titanium nitride film above the passivation layer ($SiO_2$) has etching residues between the copper lines, which is the leading cause of the reduction of the isolation resistance between the device channels. The voltage difference between the D2 pad and the copper wire promotes the ionic migration [17] of the copper element in the wiring layer and the titanium element in the barrier layer at the scratch site, thus generating a leakage channel.

This failure was due to internal scratches and incomplete etching of copper wiring arising from a specific process defect in the production of SiPM. This defect was not exposed during the stress quality assurance experiment. In the GRD calibration experiment, the energy spectrum signal measured by the SiPM array was normal, but the current increased somewhat. During the test of the whole GECAM payload, the SiPM copper wiring had a leakage channel between the signal output terminal and the power ground due to ionic migration, and the SiPM bias voltage of the GRD detector was not controlled by the triode of the data acquisition board. Table 1 shows the current records of the faulty SiPM array at different times. As it was impossible to measure the current of a single SiPM array during of the whole GECAM payload, no current data was available. After the long time aging, the defects of SiPM can be detected. The defects can make the dark current increase. We contacted the SiPM manufacturer (SensL, Ireland), and the manufacturer responded that the failure probability of newly produced SiPM devices was $1/10^9$ based on previous statistics. We finally replaced the whole failure SiPM array with a backup copy.

Table 1 Records of current status of the failure SiPM array

| Date | Current (μA) | Stage |
| --- | --- | --- |
| 20191215 | 96 | After the initial welding and assembly of the SiPM array. |
| 20200116 | 102 | After quality assurance and thermal cycling test. |
| 20200515 | 183 | Records of current in the GRD |

| | | calibration experiment. |
| --- | --- | --- |
| 20200711 | Not available | Unable to turn off the bias voltage during test of the whole payload. |
| 20200804 | 280 | Retest the GRD after confirming the SiPM array failure. |

## 4. Conclusions and discussion

In the SiPM application of GECAM, quality assurance experiments were conducted in accordance with aerospace specifications and the mechanism of the failure SiPM devices was analyzed during the development process.

A total of 79 SiPM arrays were assembled in the GECAM project. After quality assurance experiments, 50 SiPM arrrays were selected as the final products for installation on the satellite payload. However, during the joint test of the whole payload, one piece of failure SiPM appeared in 3200 SiPMs of 50 GRDs. Based on a failure analysis, it was found that a piece of SiPM had a leakage channel after longtime operation due to device defects. The etching residue on the barrier layer during the production of SiPM caused resistance decrease between the device D1 and D2 pads. Metal ions were migrating under the voltage difference when the device was working, which eventually led to formation of leakage paths. The SiPM signal output terminal and the power ground network is short-circuited because of leakage channel. Based on the quality assurance test results, the final pass rate of SiPM array was 95%.

According to the accumulated experience, in the reliability test of SiPM, it is necessary to pay special attention to test the impedance of each pin of SiPM to ground, and confirm that the power switch state of SiPM is controllable. In addition, long time tracking test is needed to ensure the normal operation of SiPM array in the long-term use. For the in-flight operation, because the power supply of 64 pixels SiPM array is divided into two parts, when a piece of SiPM fails, the corresponding part can be shut off and the residual 32 pixels still ensure the overall maintenance work. For future SiPM inflight applications, multi-voltage power supply and multi-electronic residual can improve system reliability. Heating plate can be used to warm up the SiPM and the radiation damage can partially recovered by annealing procedures[16].

Based on the SiPM application experience in GECAM, the subsequent SiPM array design will be improved. The SiPM data sheet recommends the NC pins should be directly connected to the power or ground plane. However, as there may be a small probability of short circuit between the pins after long-term use, it is necessary to connect the NC pin to an independent network and then connects to ground plane with a high-resistance resistor in future applications. In addition, in future quality assurance experiments, it is also necessary to test whether the SiPM power supply can be switched on and off normally." The GECAM satellite was launched on December 10, 2020. All SiPM arrays are working properly. Due to the space radiation effect of SiPM [18] [19], the dark current of SiPM will increase with time during in-orbit operation, so it is necessary to continuously monitor it during in-orbit operation to provide valuable monitor data for the future space experiments.

*Acknowledgements:*

*We would like to express our appreciation to the staff of the Key Laboratory of Particle Astrophysics, Center for Space Science and Applied Research and and Shandong Institute of aerospace electronic technology who offer great help in the phase of development. This research is supported by the Strategic Priority Research Program of Chinese Academy of Sciences, Grant No. XDA15360102. The authors also would thank the anonymous reviewers for their detailed and constructive comments in evaluation this paper.*